\documentclass[11pt]{article}

\usepackage[final]{acl}

\usepackage{times}
\usepackage{latexsym}

\usepackage[T1]{fontenc}

\usepackage[utf8]{inputenc}

\usepackage{microtype}

\usepackage{inconsolata}

\usepackage{graphicx}

\usepackage{graphicx}
\usepackage{amsmath}
\usepackage{comment}
\usepackage{caption}
\usepackage{subcaption}
\usepackage{wrapfig}
\usepackage{makecell}

\usepackage{url}            
\usepackage{booktabs}       
\usepackage{amsfonts}       
\usepackage{nicefrac}       
\usepackage{microtype}      
\usepackage{xcolor}         
\usepackage{graphicx}
\usepackage{amsmath}
\usepackage{float}
\usepackage{appendix}

\usepackage{xcolor}
\usepackage{fontawesome5}

\definecolor{mypurple}{RGB}{128,0,128}

\usepackage{hyperref} 

\title{Non-invasive electromyographic speech neuroprosthesis: a geometric perspective}

\author{
  \textbf{Harshavardhana T. Gowda\textsuperscript{}} and 
  \textbf{Lee M. Miller\textsuperscript{}}\\
\textsuperscript{}University of California, Davis
\\\textbf{Correspondence:} \href{tgharshavardhana@gmail.com}{tgharshavardhana@gmail.com}
}

\begin{document}

\maketitle

\begin{abstract}
We present a neuromuscular speech interface that translates silently voiced articulations directly into text. We record surface electromyographic (EMG) signals from multiple articulatory sites on the face and neck as participants {\em silently} articulate speech, enabling direct EMG-to-text translation. Such an interface has the potential to restore communication for individuals who have lost the ability to produce intelligible speech due to laryngectomy, neuromuscular disease, stroke, or trauma-induced damage (e.g., radiotherapy toxicity) to the speech articulators. Prior work has largely focused on mapping EMG collected during \emph{audible} articulation to time-aligned audio targets or transferring these targets to \emph{silent} EMG recordings, which inherently requires audio and limits applicability to patients who can no longer speak. In contrast, we propose an efficient representation of high-dimensional EMG signals and demonstrate direct sequence-to-sequence EMG-to-text conversion at the phonemic level without relying on time-aligned audio.\\

\textcolor{magenta}{\Large \faGlobe} \href{https://harshavardhanatg.github.io/emg2text.github.io/}{\textsc{Project Page}}. \textcolor{magenta}{\Large \faGithub} \href{https://github.com/HarshavardhanaTG/emg2speech}{\textsc{GitHub}}. \textcolor{magenta}{\Large \faDatabase} \href{https://osf.io/bgh7t/files/box}{\textsc{Data}}.
\end{abstract}

\section{Introduction}
Electromyographic (EMG) signals collected from the orofacial neuromuscular system during the silent articulation of speech in an alaryngeal manner can be synthesized into personalized audible speech, potentially enabling individuals without vocal function to communicate naturally. Moreover, such systems could seamlessly interface with virtual environments where audible communication may be disruptive (e.g., multiplayer games) or to facilitate telephonic conversations in noisy settings. A key enabler of these advancements is the rich information encoded in EMG signals recorded from multiple spatially distributed locations, capturing muscle activation patterns across different muscles. This richness allows for the decoding of subtle and intricate articulatory details, potentially offering higher bandwidth and lower latency compared to exocentric or allocentric modalities, such as video-based lip-to-speech synthesis. By leveraging this information, EMG-based systems offer a promising foundation for natural and efficient communication across a range of applications.

The works in \citet{willett2023high}, \citet{card2024accurate}, and \citet{metzger2023high} present invasive speech brain-computer interfaces (BCI). While invasive methods are suitable for individuals with complete anarthria, e.g., due to advanced amyotrophic lateral sclerosis, our EMG-based non-invasive speech prosthesis is appropriate for individuals with a broad range of speech impairments including dysarthria and dysphonia/aphonia, e.g., in those who have undergone laryngectomy. Work in \citet{defossez2023decoding} demonstrates a non-invasive BCI in which listened speech segments are reconstructed from magnetoencephalography (MEG) or electroencephalography (EEG) signals. However, such systems are not suitable for initiating communication (e.g., through speech).

Unlike invasive methods \citep{willett2023high, card2024accurate, metzger2023high}, which can record neural activity at single-neuron resolution with high signal-to-noise ratios, EMG captures the aggregated activity of multiple muscle motor units, with signals further distorted as they propagate through the subcutaneous tissue and skin. These distortions lead to spatial signal correlations across electrodes, where activity at one sensor can influence measurements at others. To model this structure and to capture patterns across different muscles, we introduce symmetric positive definite (SPD) matrix representations that encode second-order inter-channel correlations, providing a compact and discriminative representation of EMG signals. In contrast to prior approaches  \citep{defossez2023decoding, gaddy2020digital, gaddy2021improved}, which learn representations by mapping time-aligned MEG, EEG, or EMG signals to corresponding audio, we further improve the translation pipeline by directly predicting phoneme sequences from EMG without requiring time-aligned audio. This is achieved using connectionist temporal classification (CTC) loss \citep{graves2006connectionist}, enabling alignment-free sequence prediction akin to standard speech-to-text (S2T) translation.
\section{Prior work}

The current benchmark for silent speech interfaces is established by
\citet{gaddy2020digital, gaddy2021improved}.
In these works, electromyographic (EMG) signals recorded during
\emph{silently} articulated speech ($E_S$) and \emph{audibly} articulated speech
($E_A$), together with the corresponding audio ($A$), are used to
train recurrent neural transduction models.
These models learn a mapping from \emph{time-aligned} EMG features
($E_A$ or $E_S$) to audio ($A$).
In the baseline formulation, joint representations between $E_A$ and $A$ are
learned during training and subsequently evaluated on $E_S$.
An improved variant further aligns $E_S$ with $E_A$ and uses the aligned features
to strengthen the learned EMG-audio representation.

Despite strong performance, these approaches have several fundamental
limitations that restrict their applicability in real-world clinical settings.
Specifically, they require:
\textcircled{\footnotesize 1} access to high-quality $E_A$ and audio $A$, which
may be unavailable or unreliable in individuals with impaired
articulation, such as laryngectomy (absence of laryngeal voicing) or ALS (degraded acoustic recordings due to bulbar impairment);
\textcircled{\footnotesize 2} the need for a \emph{2x} sized training corpus for learning {\em x} representations (requiring both $E_A$ and $E_S$);
and \textcircled{\footnotesize 3} accurate temporal alignment between EMG and
audio streams, which is computationally expensive and difficult to obtain
robustly, thereby limiting scalability and near real-time deployment.
In contrast, our approach eliminates these dependencies entirely by training
directly on $E_S$ paired only with phonemic transcriptions, without any
EMG-audio alignment, using the CTC objective.

A geometric perspective on EMG representation is introduced in
\citet{gowda2024geometry}.
That work shows that, unlike images or audio signals, which are functions
sampled on Euclidean grids, multichannel EMG signals are more naturally modeled
through covariance structure, whose intrinsic geometry lies on the manifold of
symmetric positive-definite (SPD) matrices.
While \citet{gowda2024geometry} focus primarily on classification of isolated
articulatory gestures or phoneme segments, we extend this framework to
sequence-to-sequence EMG-to-phoneme modeling, enabling continuous speech
decoding. We present a detailed literature review and broad comparisons with other brain-computer interfaces (BCI) in appendix \ref{apd:lit}.

\subsection{Our contribution}

We make two primary contributions.

\textbf{First}, we open-source one of the largest high-quality EMG-to-speech
datasets collected during silent speech articulation ($E_S$).
The dataset comprises approximately 8 hours of EMG speech data from a healthy
participant, covering a large-vocabulary corpus with over 6500 unique words.
To the best of our knowledge, it is among the most comprehensive publicly
available resources for EMG-to-speech research to date.

\textbf{Second}, we demonstrate that symmetric positive definite (SPD) matrices
provide a natural and sufficient \emph{spatial representation} of EMG for
EMG-to-text decoding.
Motivated by the physiological view of EMG as arising from the additive
superposition of motor unit action potentials \citep{farina2014extraction}, we use
SPD matrices to model the spatial structure of multichannel muscle activity.
Temporal dynamics are then captured using a simple vanilla GRU operating on
eigenvalue-based representations of the SPD matrices, followed by CTC loss.
This straightforward architecture aligns with modern speech-to-text modeling
paradigms while remaining physiologically interpretable and enabling robust
phoneme-by-phoneme decoding.

Our system is trained using only silently articulated EMG signals and their
corresponding text transcriptions, and performs phoneme-level decoding directly
from EMG followed by phoneme-to-word transcription.
Unlike prior EMG-to-speech systems
\citep{gaddy2020digital, gaddy2021improved, benster2024cross}, our approach does
not assume access to time-aligned EMG-audio pairs at any stage.
Crucially, these results provide evidence that linguistically meaningful
speech structure can be inferred from muscle activity alone and transcribed into
words using only EMG.

Because our setting targets unaligned EMG-to-text generation without parallel
audio supervision, there are no existing benchmarks that enable direct
one-to-one comparisons.
Nevertheless, we compare our methods against baselines from
\textsc{emg2qwerty} \citep{sivakumaremg2qwerty}.
This comparison is well-motivated because both tasks involve decoding discrete
linguistic sequences from EMG using closely related modeling and decoding
pipelines, making \textsc{emg2qwerty} a robust and widely used benchmark for
contextualizing our methods. 

\section{Methods}
\label{Methods}
EMG signals are collected by a set of sensors $\mathcal{V}$ and are functions of time $t$. A sequence of EMG signals $E_S$ corresponding to silently articulated speech, associated with audio $A$ and phonemic content $L$, is represented as $E_S = \{\textbf{f}_v(t)\}_{\forall  \text{ v }\in\mathcal{V}}$. Here, $\textbf{f}_v(t)$ denotes the EMG signal captured at a sensor node $v$ as a function of time $t$. The audio signal $A$ encodes both phonemic (lexical) content and expressive aspects of speech, such as volume, pitch, prosody, and intonation, while $L$ represents purely the phonemic content---a sequence of phonemes. For instance, the phonemic content $L$ of the word \textsc{$<$friday$>$} is denoted by the phoneme sequence \textsc{$<$f-r-iy-d-ay$>$}.

To model the mapping from $E_S$ to $L$, we employ a sequence-to-sequence model trained using CTC loss. This approach allows us to train the model with \textit{unaligned} pairs of $E_S$ and $L$, eliminating the need for precise alignment between the input signals and their corresponding phoneme sequences. During testing, a sample of $E_S$ not in the training set outputs probabilities over all possible phonemes (40 of them in our case) at every time step, and we construct $L$ using beam search. $L$ is then converted to words using a language model.

\begin{figure*}[ht!]
  \centering
  \includegraphics[width=\textwidth]{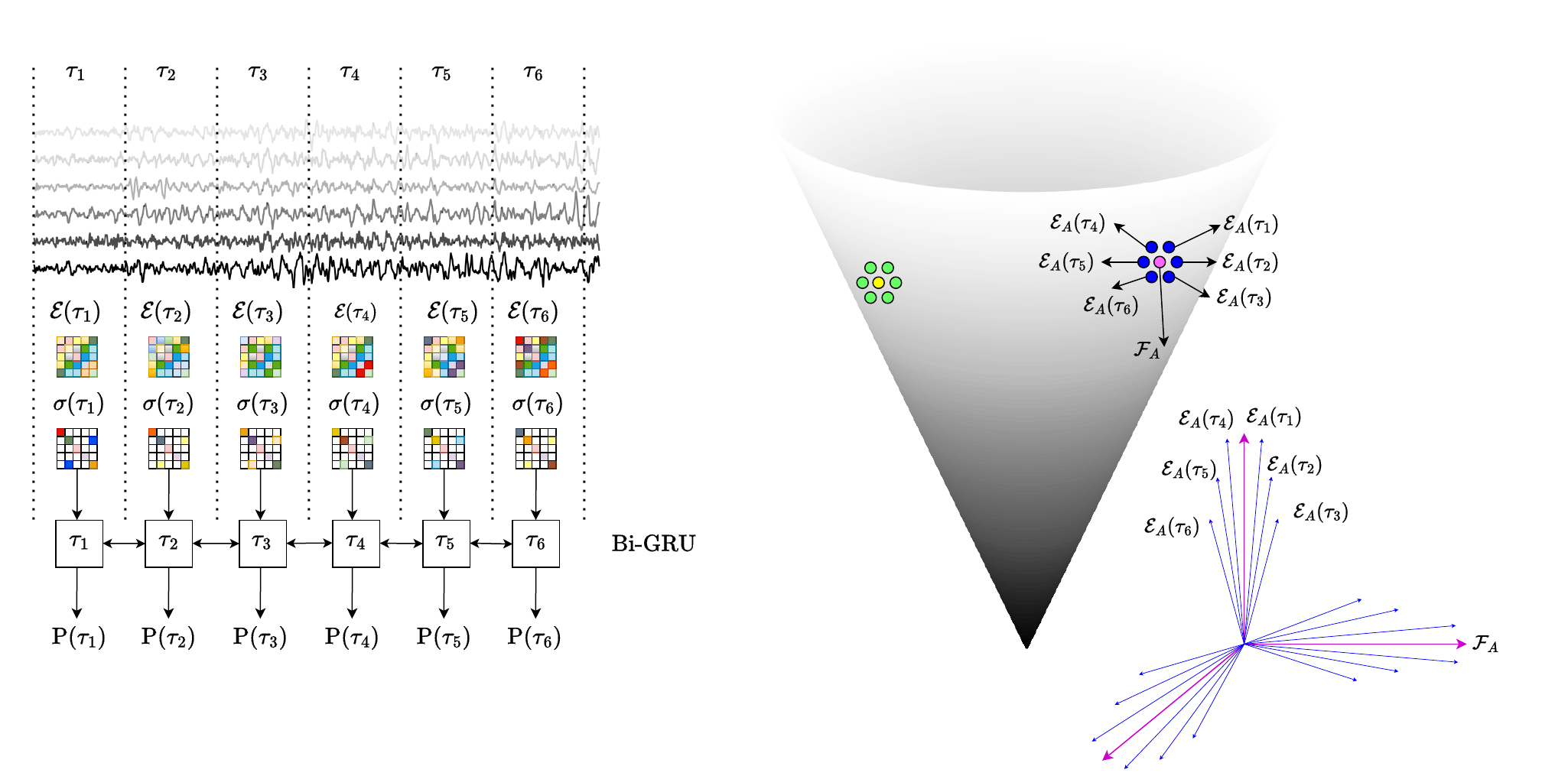}
  \caption{\textsc{Left:} EMG-to-phoneme translation pipeline. Bandpass-filtered and $z$-normalized EMG signals are converted into SPD edge matrices $\mathcal{E}(\tau)$, which are approximately diagonalized to $\sigma(\tau)$ and passed through a BiGRU. The model outputs phoneme probabilities $P(\tau)$ every 20 ms. The most probable phoneme sequence is decoded using beam search.
\textsc{Right:} Illustration of the geometry of SPD matrices in 3D. Edge matrices from individuals $A$ (blue) and $B$ (green) are shown on a convex cone manifold, with their corresponding Fréchet means in purple and yellow, respectively. The tangent spaces at $A$ and $B$ differ (because the surface is curved), and the induced transformations in $\mathbb{R}^{|\mathcal{V}|}$ reflect a change of basis. Inset: eigenvectors of individual $A$.}
  \label{fig:Illustration}
\end{figure*}
\subsection{EMG data representation}
\citet{gowda2024geometry} demonstrate that the manifold of SPD matrices serves as an effective embedding space for EMG signals, enabling the natural distinction of different orofacial movements associated with speech articulation and all English phonemes using raw signals. We make significant improvements on their methods to perform phoneme-by-phoneme decoding as opposed to classification paradigms and demonstrate our methods on continuously articulated speech in the English language as opposed to discrete word or phoneme articulations.

We construct a complete graph $\mathcal{G} = (\mathcal{V}, \mathcal{E}(\tau))$ to represent the functional connectivity of EMG signals, where $\mathcal{E}(\tau)$ denotes the set of edges over a time window $\tau = [t_{\textsc{\small start}}, t_{\textsc{\small end}}]$. The edge weight between two nodes $v_1$ and $v_2 \in \mathcal{V}$ within this time window is defined as $e_{12} = e_{21} = \frac{1}{\tau}{\textbf{f}}_{v_1}^T {\textbf{f}}_{v_2}$, which corresponds to the covariance of the signals at those nodes during the interval. Consequently, the edge (adjacency) matrix $\mathcal{E}(\tau)$ is symmetric and positive semi-definite.  To ensure positive definiteness, we convert the semi-definite adjacency matrices to definite matrices by applying the transformation $\mathcal{E} \leftarrow (1 - \eta) \mathcal{E} + \eta \, \texttt{trace}(\mathcal{E}) \, \mathcal{I}$, where $\mathcal{I}$ is the identity matrix of the same dimension as $\mathcal{E}$. We empirically found that $\eta = 0.1$ suffices for all our data. We then model these symmetric positive definite (SPD) matrices using a Riemannian geometry approach via Cholesky decomposition, as described by \citet{lin2019riemannian}.

For any adjacency matrix $\mathcal{E}$, we can express it as $\mathcal{E} = U \Sigma U^T$, where $U$ is the matrix of eigenvectors, and $\Sigma$ is a diagonal matrix containing the corresponding eigenvalues. However, instead of calculating $U$ for each $\mathcal{E}$ at every time-step $\tau$, we fix an approximate common eigenbasis $Q$ derived from the Fr\'echet mean $\mathcal{F}$ \citep{lin2019riemannian} of all adjacency matrices (at different time points) in the training set. Specifically, we compute $\mathcal{F}$ as the geometric mean of all $\mathcal{E}$, and decompose it as $\mathcal{F} = Q \Lambda Q^T$, where $Q$ contains the eigenvectors of $\mathcal{F}$, and $\Lambda$ is a diagonal matrix of its eigenvalues.

Using this fixed eigenbasis $Q$, any adjacency matrix $\mathcal{E}$ can be approximately diagonalized as $Q^T \mathcal{E}Q$, yielding a sparse matrix $\sigma$ (see figure \ref{fig:ratioLargeVocab}). This formulation allows us to work in an approximate graph spectral domain with a consistent orthogonal basis across all time windows $\tau$. For our task, we compute the graph spectral sequences $\sigma$ for all time windows $\tau$ and use these as inputs for EMG-to-language translation. We illustrate these concepts in figure \ref{fig:Illustration}.

\paragraph{Fr\'echet mean:} 
Given a set of SPD edge matrices $\mathcal{E}(\tau)$ over different time windows $\tau$, we first calculate their corresponding Cholesky decompositions $\mathcal{L}(\tau) = \textsc{cholesky}(\mathcal{E}(\tau))$, such that $\mathcal{E}(\tau) = \mathcal{L}(\tau)\mathcal{L(\tau)}^T$. Then, the Fr\'echet mean of the Cholesky decomposed matrices $\mathcal{L}(\tau)$ is given by  
\[
\begin{aligned}
\mathcal{F}_{\textsc{cholesky}}
&= \frac{1}{n} \sum_{i = 1}^{n} \lfloor \mathcal{L}(\tau_i) \rfloor \\
&\quad + \exp\left(\frac{1}{n} \sum_{i = 1}^{n} \log\!\big(\mathbb{D}(\mathcal{L}(\tau_i))\big)\right).
\end{aligned}
\]

The Fr\'echet mean $\mathcal{F}$ on the manifold of SPD matrices is calculated as 
\[
\mathcal{F} = \mathcal{F}_{\textsc{cholesky}}\mathcal{F}_{\textsc{cholesky}}^T.
\]
In the above equation, $\lfloor\mathcal{L}(\tau)\rfloor$ is the strictly lower triangular part of the matrix $\mathcal{L}(\tau)$, and $\mathbb{D}(\mathcal{L}(\tau))$ is the diagonal part of the matrix $\mathcal{L}(\tau)$.
\subsection{EMG-to-phoneme sequence translation}
We implement a gated recurrent unit (GRU) architecture for EMG-to-phoneme sequence-to-sequence modeling. The input to the GRU consists of a sequence of approximately diagonalized matrices, denoted as $\sigma$, derived over different time windows $\tau$. At each time step, the GRU model outputs probability distributions over 40 phonemes in the English language. The model is trained using CTC loss, and during inference, the most probable phoneme sequence is reconstructed using beam search decoding. The end-to-end EMG-to-language translation model is depicted in figure~\ref{fig:Illustration}. 
\subsection{Geometric perspective aligns well with biology}
We model multivariate EMG signals recorded at \( |\mathcal{V}| \) sensor nodes over different time windows \( \tau \) using symmetric edge matrices \( \mathcal{E}(\tau) \in \mathbb{R}^{|\mathcal{V}| \times |\mathcal{V}|} \), which capture pairwise relationships between sensor channels. Each matrix \( \mathcal{E}(\tau) \) can be interpreted as defining a linear transformation of the sensor space \( \mathbb{R}^{|\mathcal{V}|} \), reflecting the spatial structure of EMG activity at time \( \tau \).
This transformation admits a spectral interpretation: when \( \mathcal{E}(\tau) \) is symmetric, it can be diagonalized as
\[
\mathcal{E}(\tau) = U \Sigma(\tau) U^\top,
\]
where \( U \) is an orthonormal matrix whose columns are the eigenvectors of \( \mathcal{E}(\tau) \), and \( \Sigma(\tau) \) is a diagonal matrix of eigenvalues. In this eigenbasis coordinate system, the transformation of space is expressed as a weighted combination of the eigenvectors, with the eigenvalues in \( \Sigma(\tau) \) serving as scaling coefficients.
To reduce variability across time and to enable sequential modeling, we fix an approximate eigenbasis \( Q \in \mathbb{R}^{|\mathcal{V}| \times |\mathcal{V}|} \), and project each edge matrix into this basis:
\[
\sigma(\tau) = Q^\top \mathcal{E}(\tau) Q,
\]
yielding an approximately diagonal matrix \( \sigma(\tau) \). The diagonals of \( \sigma(\tau) \) approximate the eigenvalues of \( \mathcal{E}(\tau) \) in the shared basis \( Q \), providing a compact summary of the EMG activity at each time window. These sequences of approximate eigenvalues can then be directly modeled using a recurrent neural network to capture temporal dynamics.
This formulation aligns with the physiological origin of EMG signals: the surface EMG measurement arises from an additive superposition of motor unit action potentials, resulting in a structure that is naturally well-represented in an eigenbasis. This contrasts with modalities like speech audio, which is better modeled as time-varying filters applied to time-varying sources \citep{sivakumaremg2qwerty}.
Importantly, the eigenbasis \(Q\) is subject-specific. EMG signals from different
individuals induce different transformations \(\mathcal{E}(\tau)\) and therefore
have different eigenvectors, reflecting anatomical and physiological variability
such as subcutaneous fat thickness, muscle fiber composition, conduction
velocities, and neural drive. Consequently, distribution shifts across
individuals can be interpreted as a \emph{change of basis} in the sensor space
\(\mathbb{R}^{|\mathcal{V}|}\).

\section{Data}
We curate a large-vocabulary silent-speech EMG dataset with the number of
articulated sentences comparable to that used in
\citet{willett2023high, metzger2023high}. We adapt the language corpora from
\citet{willett2023high}, which was originally developed for a speech
brain-computer interface that translates motor-cortex neural activity into
text. Our corpus contains approximately 6500 unique words and 11000
sentences. Unlike \citet{gaddy2020digital, gaddy2021improved}, we collect only
silent-speech EMG ($E_S$), and do not collect audibly articulated EMG ($E_A$) or
audio ($A$). As a result, our task is to translate $E_S$ to language without
relying on any time-aligned $E_A$ or $A$ supervision. 

For phoneme-to-word decoding, we use a small weighted finite-state transducer (WFST) language model trained on transcripts from LibriSpeech-100 \citep{7178964}, which contain roughly 38000 sentences and 35000 unique words. See appendix \ref{apd:addExp} for additional experiments.

\subsection{Experimental details}
\label{apd:expDatails}

\begin{figure*}[t]
  \centering

  \begin{subfigure}[t]{0.32\textwidth}
    \includegraphics[width=\linewidth, trim=0 25 0 38, clip]{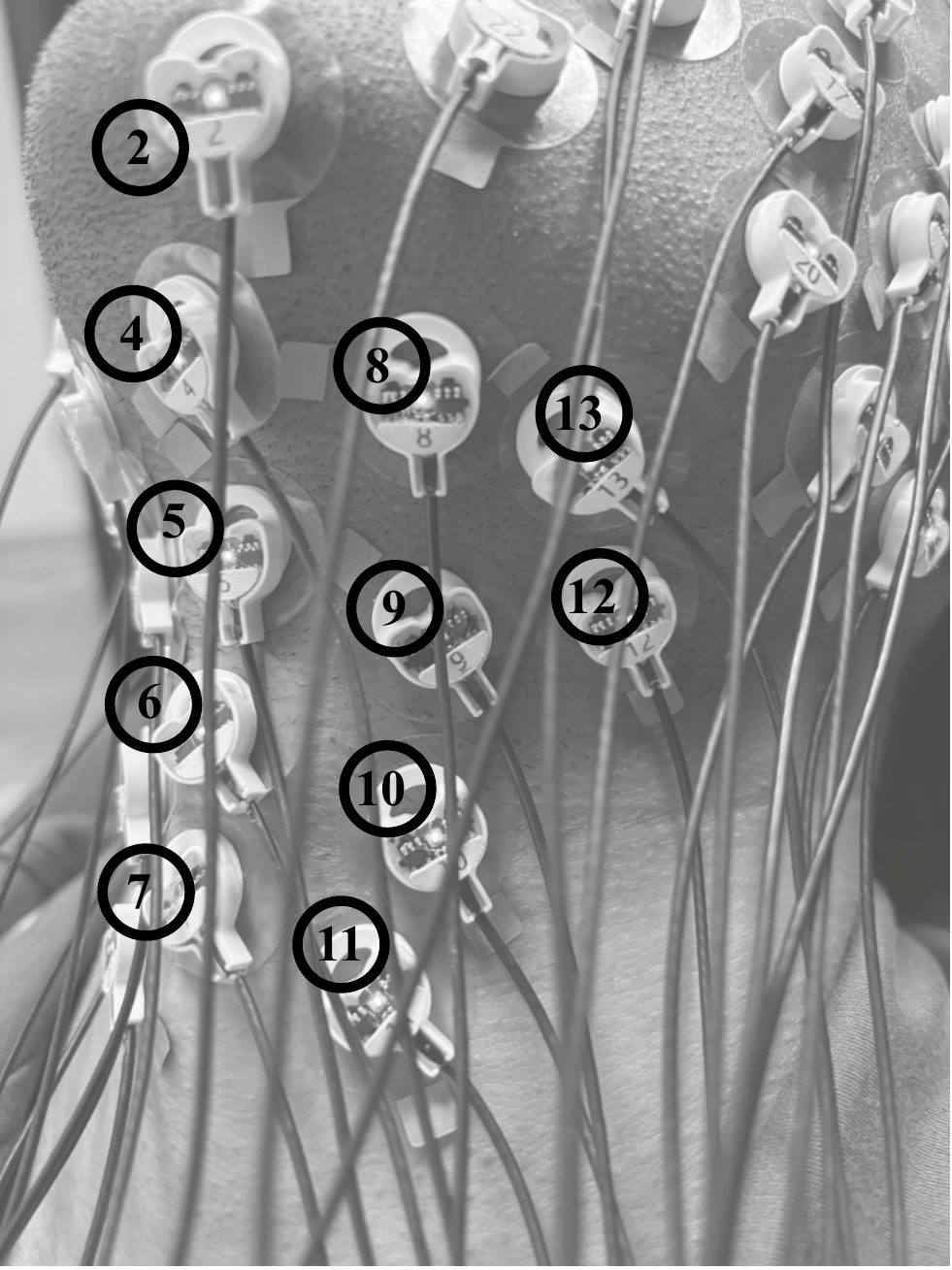}
  \end{subfigure}
  \hfill
  \begin{subfigure}[t]{0.32\textwidth}
    \includegraphics[width=\linewidth, trim=0 0 0 0, clip]{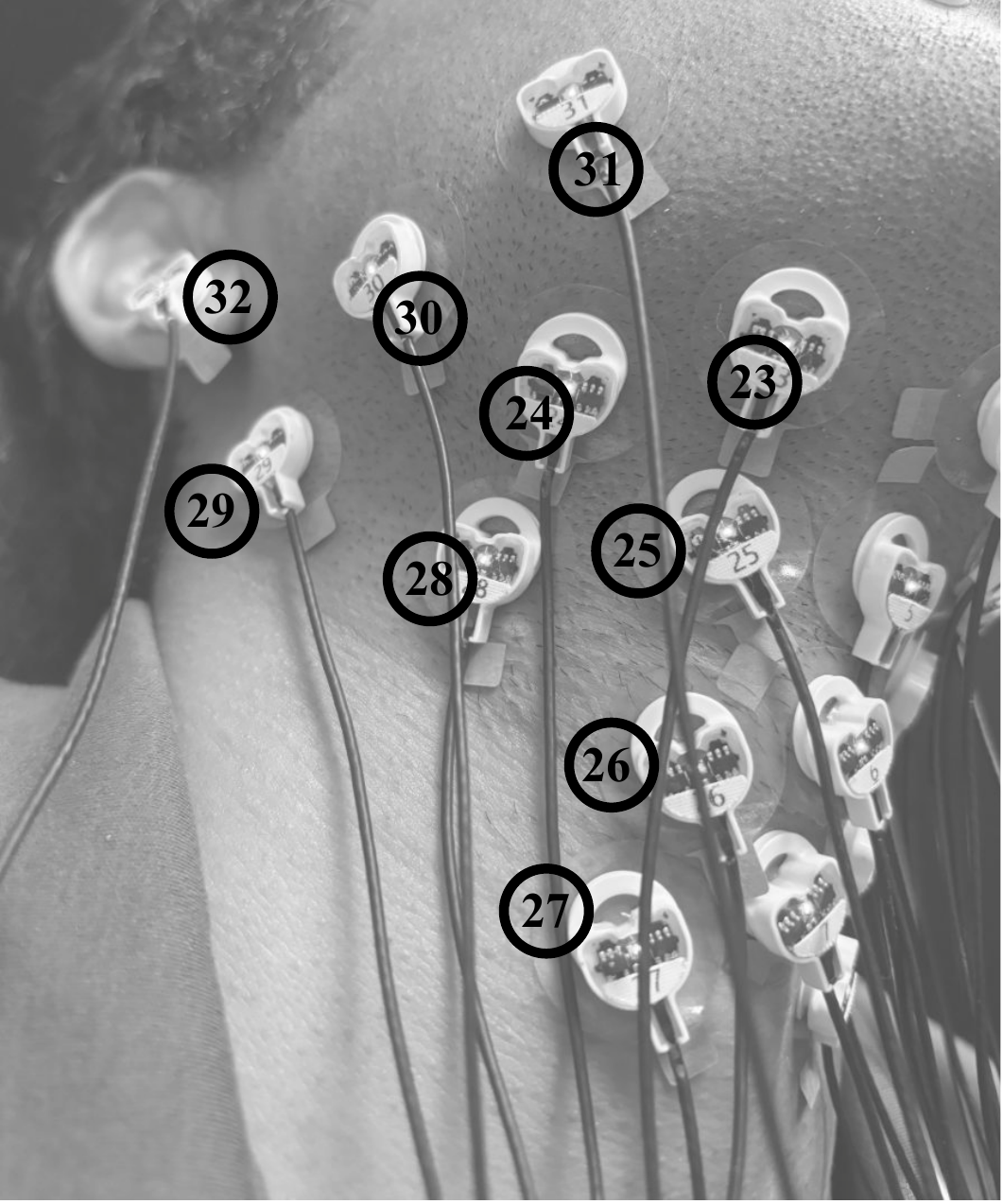}
  \end{subfigure}
  \hfill
  \begin{subfigure}[t]{0.32\textwidth}
    \includegraphics[width=\linewidth, trim=0 20 0 75, clip]{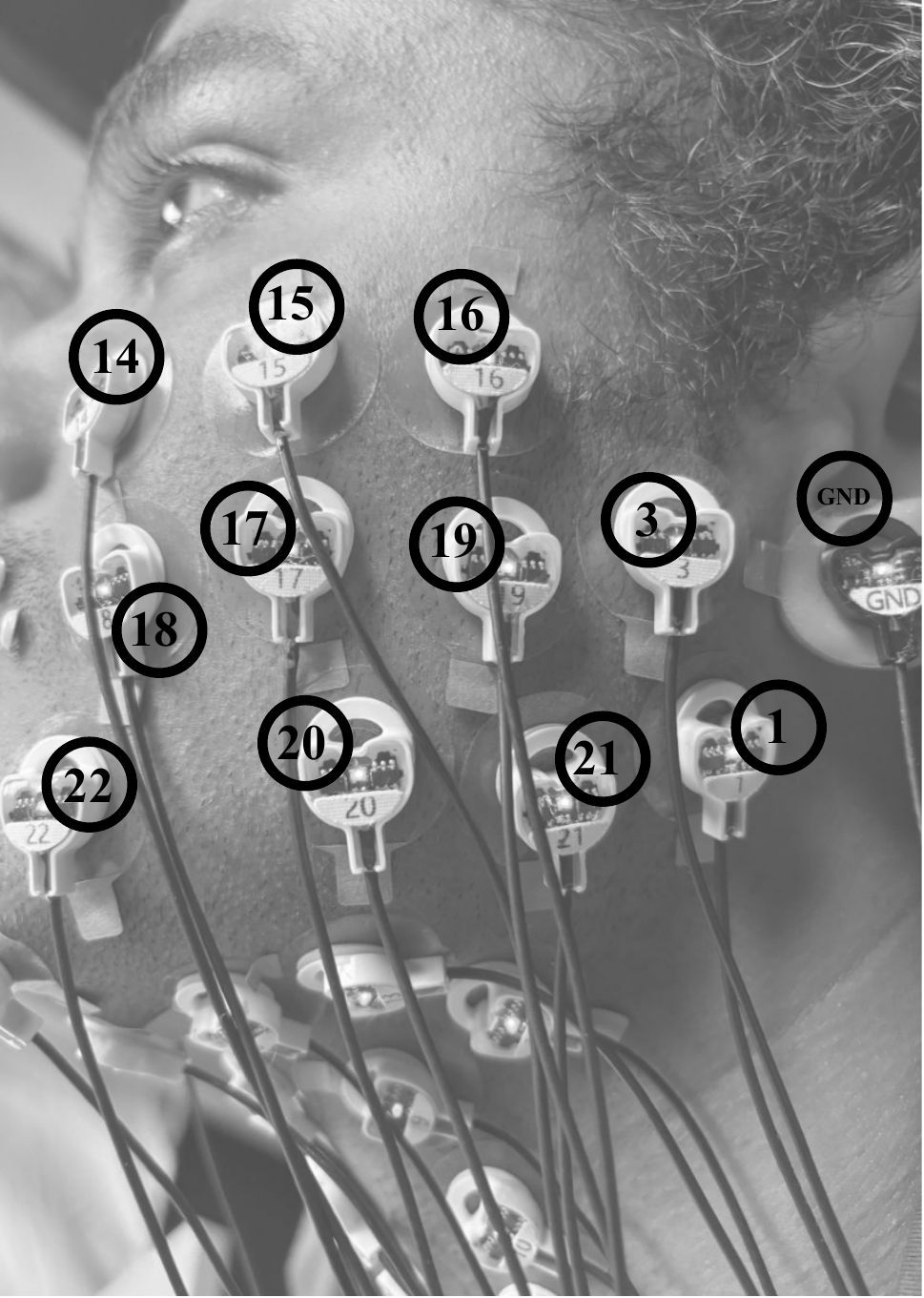}
  \end{subfigure}

  \caption{\textsc{Left:} Electrode placement on the left side of the neck. 
  \textsc{Middle:} Electrode placement on the right side of the neck. 
  \textsc{Right:} Electrode placement on the left cheek.}
  \label{fig:electrodePlacement}
\end{figure*}

We record EMG signals from 31 sites distributed across the neck, chin, jaw, cheek, and lips using monopolar electrodes. Data are acquired using an \textsc{actiCHamp Plus} amplifier with active electrodes from \textsc{Brain Vision} (\href{https://brainvision.com/products/actichamp-plus/}{Brain Vision}), sampled at 5000~Hz. To ensure low-impedance contact between the electrodes and the skin surface, we apply \textsc{SuperVisc}, a high-viscosity electrolyte gel from \textsc{Easycap} (\href{https://shop.easycap.de/products/supervisc}{Easycap}). We develop a custom software suite in a \textsc{Python} environment to present visual cues to participants and to collate and store timestamped EMG data. Time synchronization across data streams is handled using Lab Streaming Layer (LSL: \href{https://labstreaminglayer.org}{LSL}). Figure~\ref{fig:electrodePlacement} illustrates the electrode placement. In addition to the 31 data electrodes, a \textsc{ground} electrode (marked \textsc{gnd}) is placed on the left earlobe, and a \textsc{reference} electrode (marked as electrode 32) is placed on the right earlobe.

Before signal acquisition, participants are briefed on the experimental protocol and seated comfortably. For silent speech data ($E_S$), participants are instructed to articulate naturally but without producing audible speech. Sentence onset and offset are manually timestamped using mouse clicks by the participant. When ready to articulate a sentence, the participant clicks the mouse, causing the sentence to appear on the screen. After completing the articulation, the participant clicks again to mark the end of the sentence, at which point the sentence disappears from the display. This protocol allows participants to articulate each sentence at their own comfortable pace.

The data collection environment is carefully controlled to minimize AC electrical interference. EMG signals undergo minimal preprocessing. Specifically, the signal from the \textsc{reference} channel (electrode 32) is subtracted from all other channels. The resulting signals are bandpass filtered using a third-order Butterworth filter with cutoff frequencies of 80 and 1000~Hz, and segmented into individual sentences using synchronized start and end timestamps. Each segmented sentence is subsequently $z$-normalized along the time dimension on a per-channel basis. The preprocessed EMG signals are then used to construct a fully connected sensor graph, $\mathcal{E}(\tau)$, along with its approximately diagonalized representation, $\mathcal{\sigma}(\tau)$.

The electrodes are positioned over anatomical regions that directly overlie muscle groups involved in speech articulation, providing coverage of key articulators such as the tongue, jaw, lips, and larynx. Electrode locations 19, 21, 3, and 1 approximately overlie the \text{\em hyoglossus}, \text{\em palatoglossus}, and \text{\em styloglossus} muscles. These muscles, located primarily in the lower cheek and tongue regions, play a central role in tongue shaping and movement and are consistently recruited across a wide range of articulatory gestures. Muscles in the upper and posterior cheek regions—such as the \text{\em masseter} and \text{\em temporalis}, which control jaw motion, and the \text{\em zygomaticus}, which contributes to upper lip elevation—correspond approximately to electrode regions around nodes 22, 18, 17, and 15 in figure~\ref{fig:electrodePlacement}. Electrodes located beneath the jaw capture activity from muscles involved in tongue protrusion and jaw-tongue coordination, including the \text{\em genioglossus} (near electrodes 8, 9, 23, and 25) and the \text{\em digastric}. Finally, electrodes near the laryngeal region (nodes 6, 7, 10, 11, 26, and 27) reflect activity from muscles that modulate laryngeal and hyoid position—such as the \text{\em sternohyoid}, \text{\em stylohyoid}, and \text{\em digastric}—which contribute to pitch modulation, vowel shaping, and coordinated jaw movement.

\section{Results}
We use a timestep $\tau$ of 20 ms, implemented as a sliding window with 50 ms of overlapping context and a 20 ms step size, to compute $\mathcal{E}(\tau)$ and $\sigma(\tau)$, both of which are SPD matrices of size $31 \times 31$. The matrices $\sigma(\tau)$ are then input to a GRU for EMG-to-phoneme sequence translation. The dataset is split into training, validation, and test sets consisting of 8000, 1000, and 1970 sentences, respectively. Sentences in the test set are not present in the training and validation sets. The model depicted in figure \ref{fig:Illustration} is trained using 3 GRU layers for 100 epochs, and the weights corresponding to the lowest validation loss are selected.

In table~\ref{tab:largeVocab}, we report phoneme error rate (PER) and word error
rate (WER), computed as the normalized Levenshtein distance between the reference
and predicted sequences at the phoneme and word levels, respectively. To compute
WER, we convert predicted phoneme sequences into word sequences using weighted
finite-state transducer (WFST) decoding. We use transcripts from LibriSpeech-100
\citep{7178964} (approximately 38000 sentences and 35000 unique words) to
build the lexicon and language model. Following \citet{35080}\footnote{We use the WFST decoding implementation provided by \textsc{icefall}
\(\left(\raisebox{-0.2ex}{\includegraphics[height=3ex]{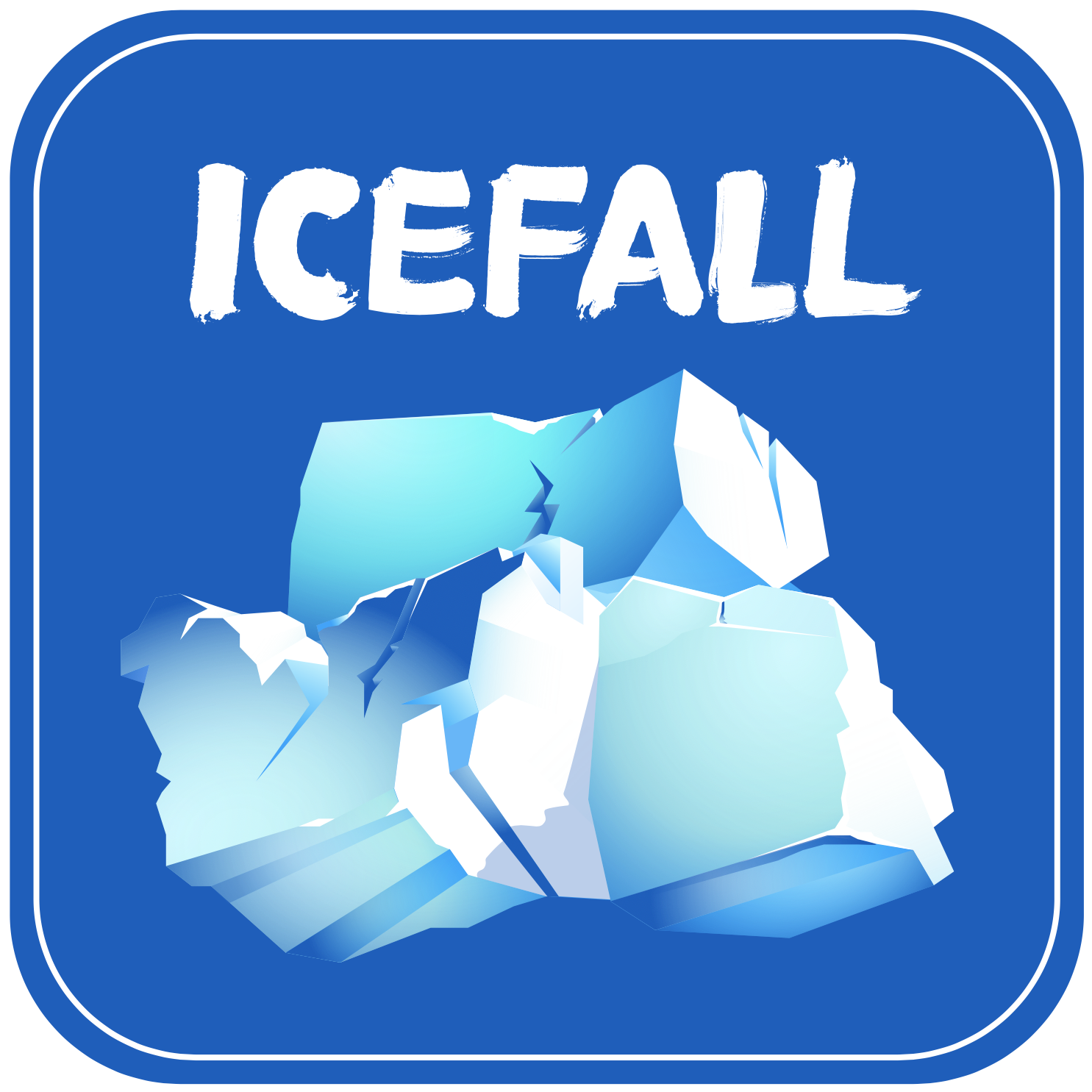}}
\href{https://github.com/k2-fsa/icefall}{\texttt{ github.com/k2-fsa/icefall}}\right).\)}, we compose the
CTC topology FST {\em H}, lexicon FST {\em L}, and an {\em n}-gram language model
FST {\em G} into a single decoding graph,
$\text{\em HLG} = \text{\em H} \circ \text{\em L} \circ \text{\em G}$.
Specifically, {\em H} encodes the allowable label sequences under the CTC
criterion, {\em L} maps phoneme sequences to word sequences, and {\em G} is
constructed from a 4-gram language model trained with KenLM
\citep{heafield-2011-kenlm}. At inference time, we perform beam search over
{\em HLG} with a beam width of 50 to obtain the best-scoring word sequence.

For comparison with prior work, we derive EMG spectrograms, in which we match the temporal resolution to that of the SPD features (50 {ms} window and 20 {ms} hop). We compute STFT (short-time Fourier transform) with $\text{\em n}_\text{FFT}$ = 256 (129 linear-frequency bins) and then average-pool the frequency axis down to 31 bins per channel. This produces per-frame tensors of shape 31 channels $\times$ 31 frequency bins, paralleling the 31 $\times$ 31 shape of $\sigma(\tau)$. 

Unlike SPD matrices $\sigma(\tau)$, which encode cross-channel articulatory structure and allow a vanilla GRU to learn meaningful temporal dependencies, raw spectrograms do not provide an equivalent inductive bias. When we use spectrograms as GRU inputs, the model collapses to predicting a small set of phoneme sequences largely independent of the input, which made phoneme-to-word decoding unreliable and resulted in a WER of $100\%$ (table~\ref{tab:largeVocab}).\footnote{One might argue that spectrogram inputs could be made more amenable to recurrent modeling through careful normalization. To enable a fair comparison with SPD matrices, we keep the decoding pipeline identical across representations: we feed either SPD matrices $\sigma(\tau)$ or raw spectrogram features into the same GRU decoder. However, on the \textsc{emg2qwerty} benchmark \citep{sivakumaremg2qwerty}, we still find that $\sigma(\tau)$ outperforms normalized spectrogram features. Our primary motivation for this comparison is to highlight that, with an SPD-matrix representation, EMG-to-text translation is feasible even with an architecture that mainly models temporal dependencies and performs no explicit spatial modeling (i.e., a vanilla GRU), because cross-channel structure is already captured by second-order (covariance) features. In contrast, spectrogram features do not appear to confer the same inductive bias under an otherwise identical recurrent decoder. This result should be interpreted in that context.
}

\begin{table}[t]
\centering
\caption{Mean PER and WER. Lower values indicate better performance.}
\begin{tabular}{lcc}
\hline
\textsc{Model} & \textsc{per}(\footnotesize$\%\downarrow$) & \textsc{wer}(\footnotesize$\%\downarrow$)\\
\hline
\makecell{\textsc{Baseline}\\\textsc{(spectrogram)}} & 89.25 & 100 \\\hline
\makecell{\textsc{Matrices $\sigma(\tau)$}\\\textsc{(ours)}} & \textbf{48.47} & \textbf{73.53} \\
\hline
\end{tabular}
\label{tab:largeVocab}
\end{table}

\begin{figure}[htbp] \centering \includegraphics[width=0.7\linewidth]{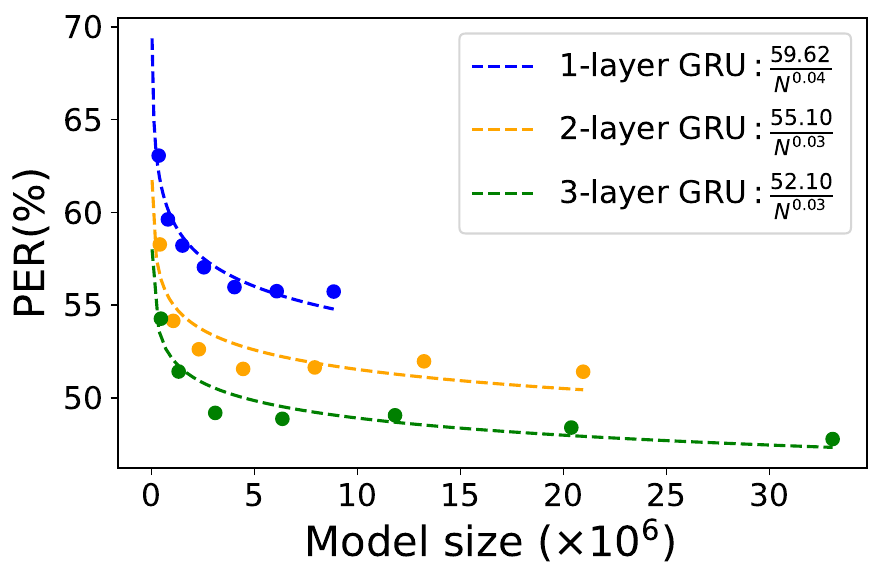} \caption{Model size versus PER for EMG-to-phoneme translation.} \label{fig:perSize} \end{figure}

In figure~\ref{fig:perSize}, we study how model capacity affects phoneme error rate (PER).
We vary the number of GRU layers and the hidden-state dimensionality, which changes the
total number of trainable parameters $N$.
Across the range of models explored, PER decreases as $N$ increases and is
approximately consistent with a power-law-like trend, $\text{\em PER} \propto N^{-\beta}$ for $\beta>0$,
similar in spirit to empirical scaling behaviors reported for neural language models
\citep{kaplan2020scaling}.
The observed trend suggests a predictable relationship between model capacity and decoding
accuracy in this setting.
Notably, even a single-layer GRU attains a PER of 56\%, with deeper and wider models
yielding further improvements.

\begin{figure}[htbp] \centering \includegraphics[width=0.7\linewidth]{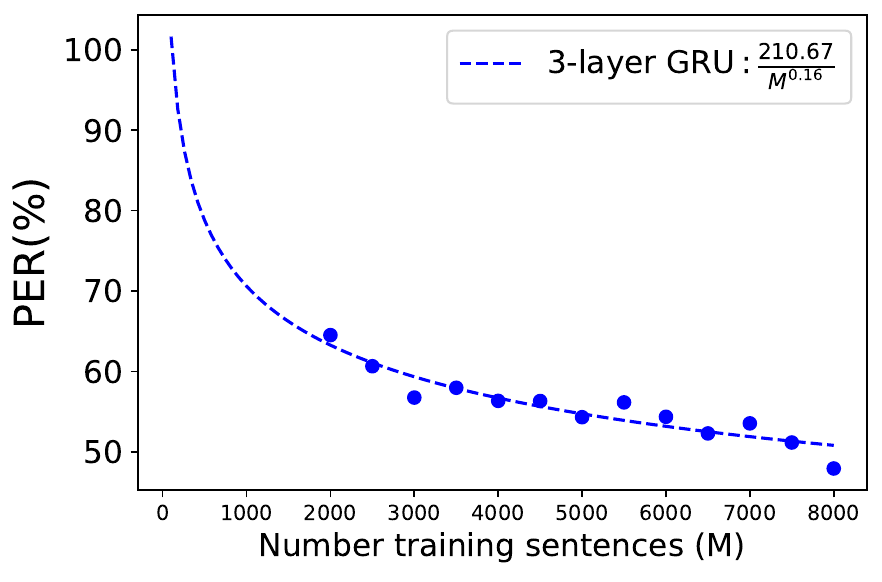} \caption{Training data size versus PER for EMG-to-phoneme translation.} \label{fig:dataSize} \end{figure}

In figure~\ref{fig:dataSize}, we examine how the amount of training data affects PER.
We vary the number of training sentences $M$ while keeping the validation and test sets fixed.
PER again decreases with more data and is approximately consistent with a
power-law-like trend over the range explored, $\text{\em PER} \propto M^{-\beta}$ for $\beta>0$
\citep{kaplan2020scaling}.
The result indicates a regular relationship between data availability and decoding accuracy
in our EMG-to-phoneme setting.
Together, these trends suggest that scaling both model capacity and data can improve
performance, with diminishing returns at larger sizes.

 


\subsection{Comparison with prior work}

\label{subsec:emg2qwerty}
\begin{table*}[htpb]
\centering
\caption{Comparison between our proposed methods and those presented by \cite{sivakumaremg2qwerty}, with all results averaged over 8 subjects. Parameter size and FLOPs are identical across all the models. Lower CER is better. The CER improvement arising from our method is statistically significant ($p < 0.015$). LM: language model.
}
\[
\begin{array}{l | c c | c c}
\hline
\textbf{} & \multicolumn{2}{c|}{\textsc{No LM}} & \multicolumn{2}{c}{\textsc{6-gram char-LM}} \\
 & \small {\textsc{Val CER ($\%\downarrow$)}} & \small{\textsc{Test CER ($\%\downarrow$)}} & \small{\textsc{Val CER ($\%\downarrow$)}} & \small{\textsc{Test CER ($\%\downarrow$)}} \\
\hline
\small{\makecell{\textsc{Baseline} \textsc{ (spectrogram)}\\\text{\citep{sivakumaremg2qwerty}}}} & 15.65 \pm 5.95& 15.38 \pm 5.88& 11.03 \pm 4.45& 9.55 \pm 5.16\\
\hline

\small{\textsc{Matrices $\sigma(\tau)$ (ours)}} & {\bf 14.33 \pm 5.27}& {\bf 14.03 \pm 5.27}& {\bf 9.61 \pm 3.84}& {\bf 7.95 \pm 4.54}\\
\hline
\end{array}
\]

\label{tab:comp}
\end{table*}

\begin{figure*}[htbp]
  \centering
  \begin{subfigure}[t]{0.48\textwidth}
    \includegraphics[width=0.75\linewidth]{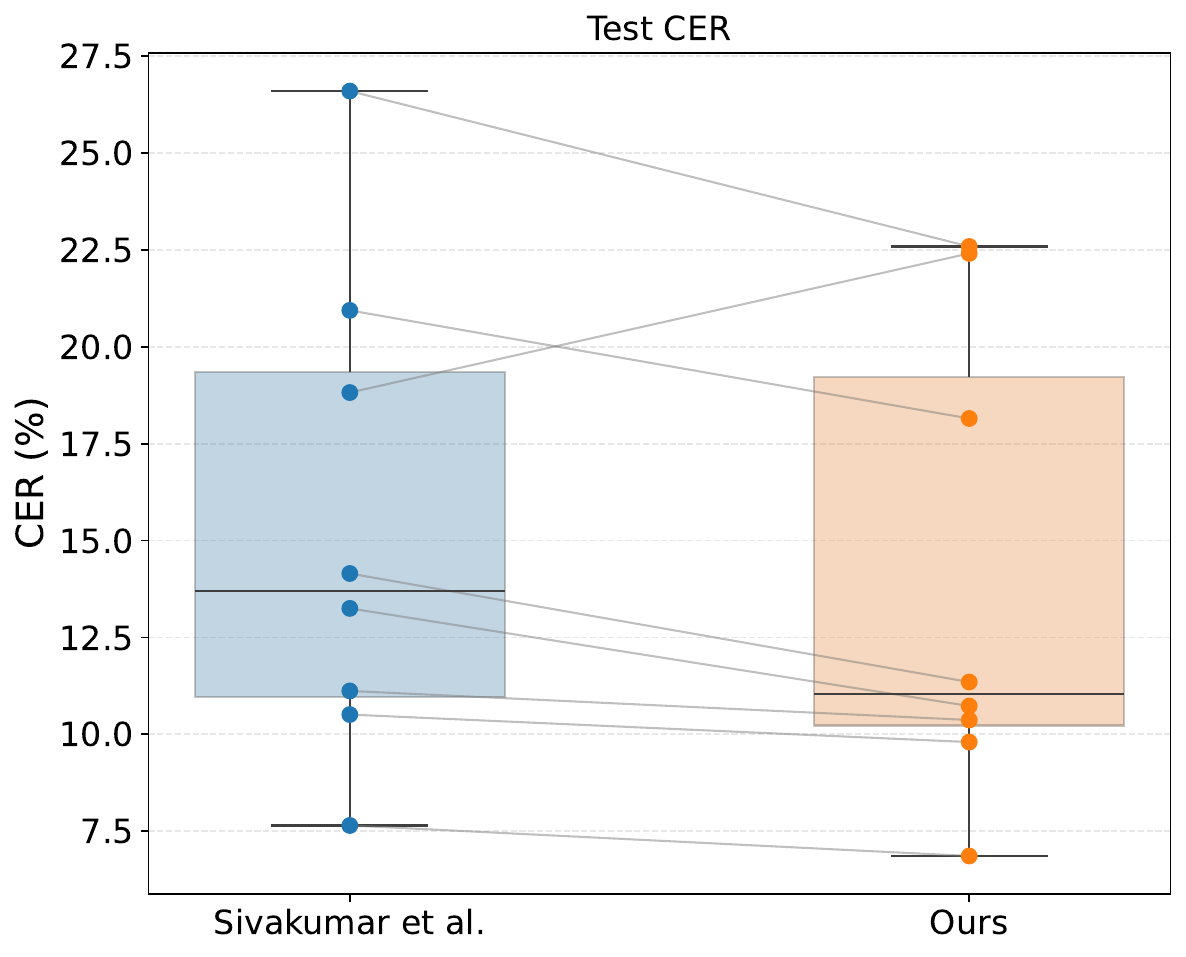}
  \end{subfigure}
  \hfill
  \begin{subfigure}[t]{0.48\textwidth}
    \includegraphics[width=0.75\linewidth]{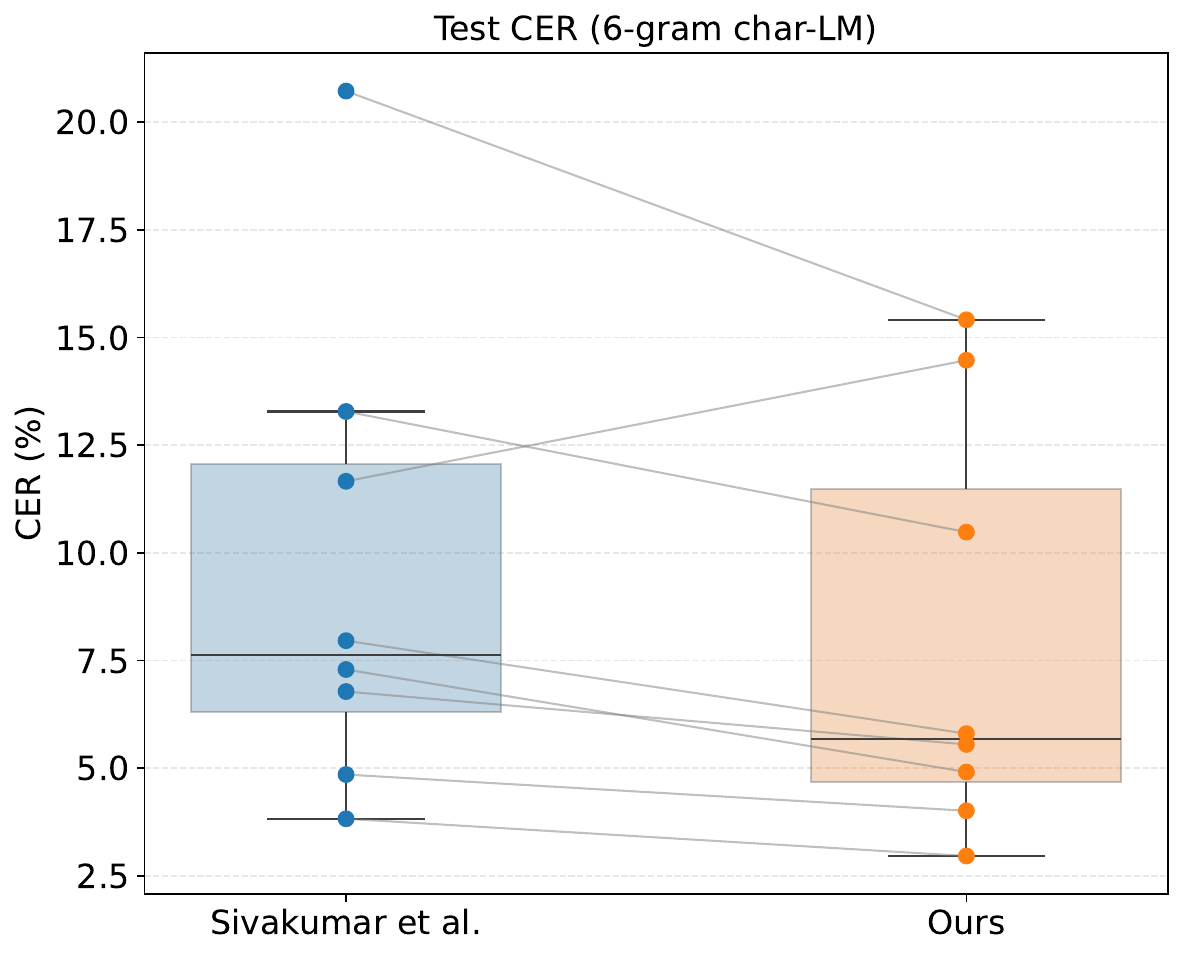}
  \end{subfigure}
  \caption{Results for individual subjects in \textsc{emg2qwerty} dataset. Each dot represents an individual test subject, with connecting lines indicating within-subject performance across different models. The boxplots summarize the median and interquartile range of the results. Our method improves performance for all subjects except \textsc{user6}.
}
  \label{fig:compQwerty}
\end{figure*}

To the best of our knowledge, there is no prior work that performs $E_S$-to-language conversion without using $E_A$ or $A$ on large English language corpora with CTC loss. Therefore, we compare our methods on the \textsc{emg2qwerty} dataset introduced by \citet{sivakumaremg2qwerty}. In this dataset, subjects wear EMG wristbands on both hands and touch-type on a QWERTY keyboard. The goal is to decode the resulting EMG signals into a sequence of characters using CTC loss. Although the physical actions involved in EMG-to-text decoding and \textsc{emg2qwerty} differ, the underlying machine learning principles remain similar.

To enable a fair comparison, we conduct controlled experiments in which we replace the original log-spectrogram features from \citet{sivakumaremg2qwerty} with SPD matrices \(\sigma(\tau)\). Apart from substituting the features, we omit their \textsc{SpecAugment} data augmentation strategy—this should not compromise the fairness of the comparison, as \textsc{SpecAugment} was shown to improve their performance. Additionally, we train our models for 250 epochs (compared to 150 in their setup, where their model converged early), and apply a weight decay of $10^{-3}$ to the Adam optimizer to ensure stable training.

We focus on a specific case from \citet{sivakumaremg2qwerty}, in which personalized models are trained independently for each individual, starting from random weight initialization. The zero-shot paradigm, in which a model is trained on data from 100 subjects and evaluated on 8 unseen individuals, as well as the personalized fine-tuning paradigm, in which individual models are initialized with generic weights pretrained on 100 subjects, are beyond the scope of this work. In this paper, we restrict our investigation to personalized models trained from scratch.

The results are presented in table~\ref{tab:comp}. As shown, our proposed methods outperform the baseline approaches reported by \citet{sivakumaremg2qwerty}, with just {\em one simple} modification. These findings support the effectiveness of our approach, which is specifically designed to reflect the underlying biological structure of EMG signals.

In figure~\ref{fig:compQwerty}, we present subject-wise character error rates (CER). Our method improves performance for all users except \textsc{user6}. When decoding is performed without a language model, we observe an 8.8\% relative improvement in CER. With a 6-gram character level language model (6-gram char-LM), the relative improvement is 16.8\%\footnote{For reference, in a personalized finetuning paradigm, \citet{sivakumaremg2qwerty} first trained a generic model on data from 100 subjects (nearly $100\times$ more data) and then finetuned it on 8 individual subjects, achieving a CER of 11.29\% without a language model and 6.95\% with a 6-gram character LM. The 6.95\% result reflects a strong performance ceiling enabled by large-scale pretraining. In comparison, our 7.95\% CER is obtained using only per-subject training (approximately $100\times$ less data) and is already close to this ceiling, highlighting the effectiveness of our approach.
}.
\section{Conclusion}
We show that continuous speech can be inferred from orofacial EMG by decoding articulations \emph{phoneme-by-phoneme}. This formulation is well matched to the physiology of speech production: phonemes are defined by place and manner of articulation, which should be expressed in coordinated patterns of orofacial muscle activity. Our approach is simple and interpretable: SPD matrices capture spatial structure in multichannel EMG, and a GRU models temporal dynamics. In our evaluation, we obtain a phoneme error rate (PER) of 49\%, substantially below the chance-level PER of approximately 98\%, providing strong evidence that direct EMG-to-text transcription is feasible. Although word error rate (WER) remains modest and experiments are currently limited to a single subject, these results establish a concrete baseline and motivate future work on improved modeling, stronger decoding, and broader validation across participants.

\newpage
\section{Limitations}
This work primarily focuses on demonstrating that linguistic content can be decoded from EMG signals alone, phoneme-by-phoneme, and then reconstructed into words. To our knowledge, this has not previously been shown in the context of general English corpora using methodologies widely adopted in modern speech-to-text paradigms. We view this as a necessary milestone in a longer-term research effort on which future advances can build; however, several limitations remain.

First, our current model uses a bidirectional GRU, which requires access to the full sentence before decoding and therefore does not support streaming, low-latency use. In follow-up work (\textcolor{magenta}{\faGlobe}~\href{https://harshavardhanatg.github.io/emg2speech.github.io/}{\texttt{emg2speech}}), we address this limitation by using causal models that rely only on local past context and by directly converting EMG sequences to speech.

Second, this study is demonstrated on a single healthy participant, so the results do not yet establish robustness across individuals or clinical populations. In a follow-up study (\textcolor{magenta}{\faGlobe}~\href{https://harshavardhanatg.github.io/emg2speech.github.io/}{\texttt{emg2speech}}), we extend the same overall approach to an individual with amyotrophic lateral sclerosis (ALS).

Third, we do not evaluate sustained, long-term performance of this non-invasive neuroprosthesis across days or months, including the effects of electrode shifts, fatigue, and other sources of day-to-day variability. In contrast, prior work on invasive neuroprostheses has reported stability over extended periods in related decoding settings, including brain-to-text \citep{fan2023plug} and cursor-based brain-computer interfaces \citep{wilson2025long}.

Finally, we do not explore whether large-scale pretrained EMG models can improve decoding performance or reduce the amount of subject-specific data required. Related work on EMG-based keyboard typing (\textsc{emg2qwerty}) suggests that pretraining on data from many individuals can improve accuracy after fine-tuning, although zero-shot performance remains limited \citep{sivakumaremg2qwerty}. Speech is likely more challenging than discrete key typing, and future work should investigate how to build and effectively leverage large-scale pretrained models for EMG-to-speech translation.

We are actively addressing these limitations through ongoing longitudinal studies and by expanding data collection to build larger EMG-to-speech corpora from individuals with diverse clinical etiologies, including ALS and laryngectomy.
\section{Ethical considerations}
\label{sec:ethics}
Research was conducted in accordance with the principles embodied in the Declaration of Helsinki and in accordance with the University of California, Davis Institutional Review Board (IRB) Administration protocol 2078695-1. All participants provided written informed consent. All participants also provided consent for publication of deidentified data. Volunteers of any gender and from all racial and ethnic groups were eligible to participate. Participants were required to be at least 18 years old, able to understand spoken and written English, and able to follow task instructions. Participants had no skin conditions or wounds at electrode placement sites and were excluded if they had uncorrected vision problems. Children, individuals unable to provide informed consent, and prisoners were not included in the experiments. All participants were compensated in accordance with IRB protocols.

\section*{Acknowledgments}

This work was supported by awards to Lee M. Miller from: Accenture, through the Accenture Labs Digital Experiences group; CITRIS and the Banatao Institute at the University of California; the University of California Davis School of Medicine (Cultivating Team Science Award); the University of California Davis Academic Senate; a UC Davis Science Translation and Innovative Research (STAIR) Grant; and the Child Family Fund for the Center for Mind and Brain. 

Harshavardhana T. Gowda is supported by Neuralstorm Fellowship, NSF NRT Award No. 2152260 and Ellis Fund
administered by the University of California, Davis.

\section*{Conflict of interest}

H. T. Gowda and L. M. Miller are inventors on intellectual property related to {\em silent} speech owned by the Regents of University of California, not presently licensed.

\section*{Author contributions}
\begin{itemize}
    \item Harshavardhana T. Gowda: Conceptualization, mathematical formulation, method development, data analysis, experimental design, data collection software development, data collection, and manuscript preparation.
    \item Lee M. Miller: Conceptualization, funding, and manuscript review.
\end{itemize}


\appendix
\section{Detailed literature review}
\label{apd:lit}
Here, we review prior work on speech neural and neuromuscular interfaces and contextualize our results relative to state-of-the-art methods.
A substantial body of research \cite{jou2006towards, kapur2020non, meltzner2018development, toth09_interspeech, 8114359, 8578038, littlejohn2025streaming} has laid the groundwork for EMG-based speech interfaces.
Among the earliest studies, \citet{jou2006towards} demonstrate EMG-to-speech conversion on a small corpus of 50 sentences.
\citet{kapur2020non} use a corpus of 15 sentences and, rather than performing phoneme-level decoding, formulate the task as a 15-way classification problem.
\citet{meltzner2018development} study EMG-to-text recognition for isolated words, phrases drawn from a $\sim$200-word vocabulary, and continuous sentences using a custom grammar-based recognition model over a set of 1200 scripted phrases.
\citet{toth09_interspeech} present EMG-to-speech conversion on a corpus of 500 sentences.
\citet{8114359} demonstrate EMG-to-speech conversion using up to two hours of data and 2000 utterances.

Overall, these studies rely on private datasets and task-specific pipelines, and they typically evaluate on small, constrained corpora.
In addition, the works do not release full implementations (e.g., code repositories) or sufficient methodological details to enable direct reproducibility.
As a result, it is difficult to directly compare performance across systems, and all the above results do not establish generalization to open-vocabulary English settings. 

A reproducible benchmark for open-vocabulary EMG-to-speech conversion was introduced by \citet{gaddy2020digital, gaddy2021improved}. However, these works rely on time-aligned EMG-audio pairs for training. Building on \citet{gaddy2021improved}, \citet{benster2024cross} propose an approach that leverages an audio-only corpus in addition to paired EMG-audio data. {\em While effective in the benchmark setting}, such methods cannot be deployed in clinical scenarios where parallel EMG-audio recordings may be unavailable or unreliable. On the large-vocabulary corpus, \citet{gaddy2020digital} report a word error rate (WER) of 68\%, and \citet{gaddy2021improved} reduce this to 42\%. \citet{littlejohn2025streaming} report a WER of 74\% on the \citet{gaddy2020digital} dataset using a CNN+RNN transducer model; however, their train-test splits and implementation details are not publicly available, which prevents direct comparison.
In our setting, we address a harder learning problem by not assuming time-aligned EMG-audio pairs during training, and we report a WER of 73\% on an open-vocabulary corpus. We emphasize that these WER values should not be compared one-to-one across studies, since the data collection setup, training targets and alignment assumptions, problem formulation, and evaluation methodology differ substantially.
We report these results to provide context relative to prior EMG-based speech interfaces.

To address these limitations, we build on widely used methods in speech-to-text (S2T) domain by adapting them to the EMG setting through principled, articulatorily motivated design choices. 

Previous work by \citet{gowda2024topology} demonstrated the effectiveness of SPD matrices in decoding \textit{discrete} hand gestures from EMG signals collected from the upper limb. Furthermore, SPD matrix representations have been extensively utilized to model electroencephalogram (EEG) signals, although they have never been applied to complex tasks such as sequence-to-sequence speech decoding. For example, \citet{barachant2011multiclass, 10.1016/j.neucom.2012.12.039} employed Riemannian geometry frameworks for classification tasks in EEG-based brain-computer interfaces, while \citet{10.5555/3454287.3454945} developed regression models based on Riemannian geometry for biomarker exploration using EEG data.

The novelty of our work lies in the algebraic interpretation of manifold-valued data through linear transformations, and the development of models for complex sequence-to-sequence tasks. This approach moves beyond the conventional applications of classification and regression.

\section{Additional results}
\label{sed:addRes}
\begin{figure}[ht]
    \centering
    \includegraphics[width=0.4\textwidth]{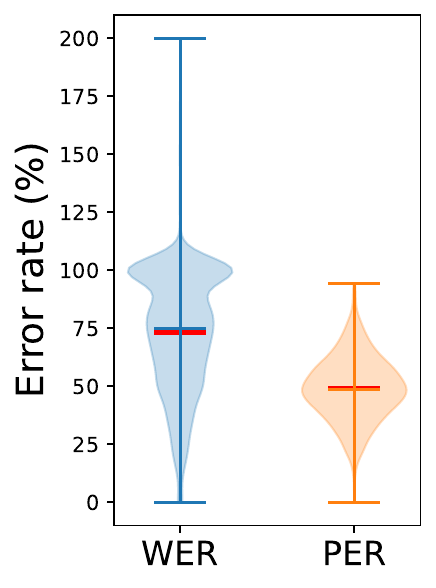}
    \caption{Distributions of WER and PER across all 1970 sentences in the test set. Means are shown in red.}
    \label{fig:werPerViolin}
\end{figure}
In figure~\ref{fig:werPerViolin}, we summarize the distribution of WER and PER across all 1970 sentences in the test set. The mean PER is 48.47\%, far below the chance-level PER of approximately \(1 - \frac{1}{40} = 97.5\%\) under uniform random guessing over 40 phoneme labels. The mean WER is 73.53\%, and the gap between WER and PER suggests that, even when word-level transcriptions are incorrect, the predicted sequences often remain phonetically plausible.
We show qualitative EMG-to-text transcription examples in table~\ref{tab:transcriptionEx}.

\begin{figure}[h!]
    \centering
    \includegraphics[width=0.48\textwidth]{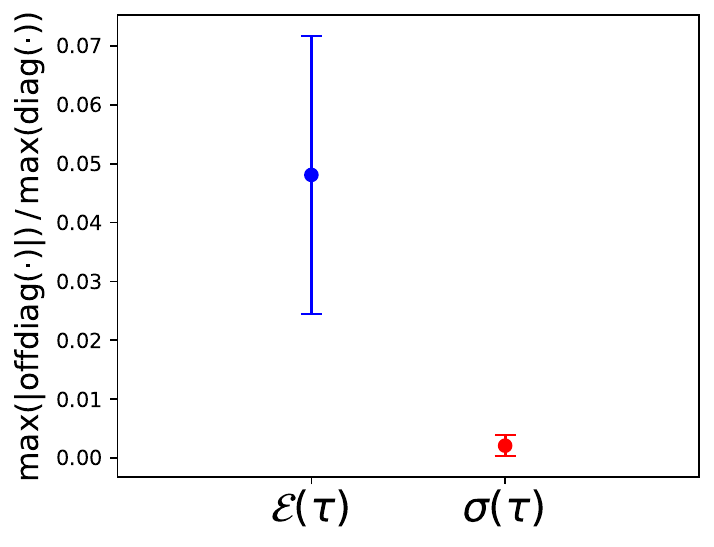}
    \caption{
    \textcolor{blue}{Blue}: dataset average of
    $\frac{\max\!\big(\lvert\operatorname{offdiag}(\mathcal{E}(\tau))\rvert\big)}{\max\!\big(\operatorname{diag}(\mathcal{E}(\tau))\big)}$
    over all $\tau$ in the train, validation, and test sets.
    \textcolor{red}{Red}: dataset average of
    $\frac{\max\!\big(\lvert\operatorname{offdiag}(\sigma(\tau))\rvert\big)}{\max\!\big(\operatorname{diag}(\sigma(\tau))\big)}$
    over all $\tau$ in the train, validation, and test sets.
    The consistently lower ratio for $\sigma(\tau)$ indicates that it is closer to diagonal (and thus sparser) than $\mathcal{E}(\tau)$.
    We use the sparse SPD matrices $\sigma(\tau)$ for EMG-to-text translation.}
    \label{fig:ratioLargeVocab}
\end{figure}

\textbf{$\sigma(\tau)$ are sparse matrices:} in figure~\ref{fig:ratioLargeVocab}, we show that $\sigma(\tau)$ is sparser than $\mathcal{E}(\tau)$; its off-diagonal entries are small relative to its diagonal, i.e., $\sigma(\tau)$ is closer to a diagonal matrix than $\mathcal{E}(\tau)$.

\section{Additional experiments}
\label{apd:addExp}
Here, we present additional experiments on small-vocabulary data and on recordings from multiple subjects. The motivation for using small-vocabulary data is to test whether we can achieve high decoding accuracy in a closed-vocabulary setting, supporting a minimum-viable neuroprosthesis. The motivation for evaluating multiple subjects is to assess whether our approach generalizes across individuals.

We curate \textsc{Data}$_{\textsc{ small-vocab}}$, a timestamped dataset of isolated and connected words, and \textsc{Data}$_{\textsc{ nato-words}}$, a compact codeword dataset based on the NATO phonetic alphabet that enables training a generalizable language-to-spelling model with minimal data.

\subsection{\textsc{Data}$_{\textsc{ small-vocab}}$}
We curate a limited-vocabulary dataset consisting of 67 unique words. These words include weekdays, ordinal dates, months, and years. Sentences are constructed in the format \textsc{$<$weekday-month-date-year$>$}. A single participant silently articulated 500 such sentences, and the resulting EMG data, denoted as {\em E$_S$}, are translated into output phoneme sequences. We have timestamps that mark the beginning and end of each word (or grouped words) within a sentence. We record EMG from 31 electrode sites at a sampling rate of 5000~Hz. For details about electrode placement and the experimental setup, please refer to section~\ref{apd:expDatails}. The sentences were presented as individual words (or grouped words), demarcated by timestamps, and displayed as follows:
\[
\underset{t=0}{<}\textsc{weekday}\underset{t=2\,\mathrm{s}}{>} \;-\;
\underset{t=2\,\mathrm{s}}{<}\textsc{month}\underset{t=4\,\mathrm{s}}{>}
\]
\[
\underset{t=4\,\mathrm{s}}{<}\textsc{date}\underset{t=6\,\mathrm{s}}{>} \;-\;
\underset{t=6\,\mathrm{s}}{<}\textsc{year}\underset{t=9\,\mathrm{s}}{>},
\]
where each segment occurs sequentially in time.\\
\\
\textbf{Results:} we use a timestep $\tau$ of 50 ms, implemented as a sliding window with 100 ms of overlapping context and a 50 ms step size, to compute $\mathcal{E}(\tau)$ and $\sigma(\tau)$, both of which are SPD matrices of size $31 \times 31$. The matrices $\sigma(\tau)$ are then input to a GRU for EMG-to-phoneme sequence translation. The dataset is split into training, validation, and test sets consisting of 370, 30, and 100 sentences, respectively. The model depicted in figure \ref{fig:Illustration} is trained using a single GRU layer for 100 epochs, and the weights corresponding to the lowest validation loss are selected.

In table \ref{tab:smallVocab}, we report the phoneme error rate (PER) and word error rate (WER), computed using the Levenshtein distance between the original and reconstructed sequences. Words are reconstructed from phoneme sequences by matching them to the word sequence with the lowest Levenshtein distance in a 67-word corpus.

\begin{table}[h!]
\centering
\caption{Mean PER and WER for \textsc{Data}$_{\textsc{ small-vocab}}$. Lower values indicate better performance.}
\begin{tabular}{cc}
\hline
\textsc{per}(\footnotesize$\%\downarrow$) & \textsc{wer}(\footnotesize$\%\downarrow$)\\
\hline
 13 & 14 \\
\hline
\end{tabular}
\label{tab:smallVocab}
\end{table}

\begin{figure}[htbp]
  \centering
  \includegraphics[width=0.8\linewidth]{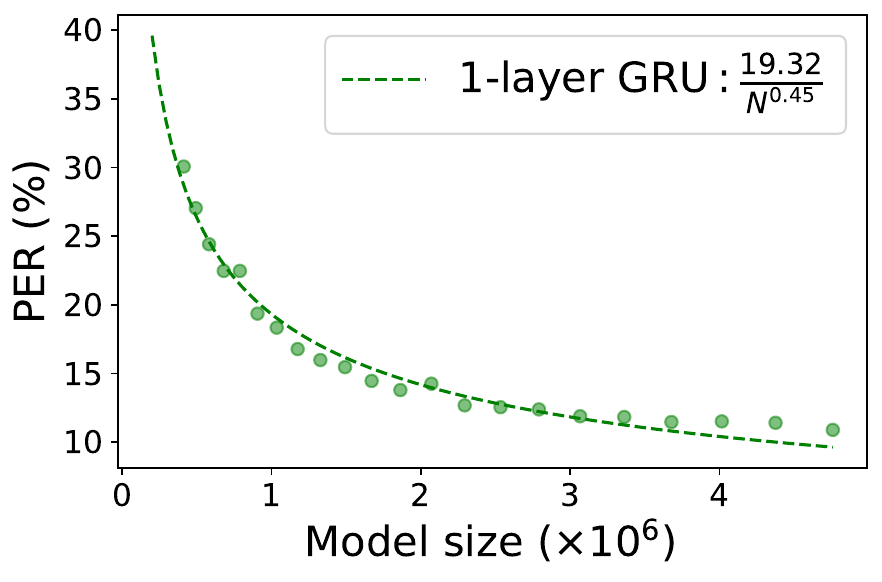}  
  \caption{Model size versus PER for EMG-to-phoneme translation for \textsc{Data}$_ \textsc{ small-vocab}$.}
  \label{fig:perSmall}
\end{figure}

In figure~\ref{fig:perSmall}, we analyze the impact of model size on phoneme error rate (PER) across different GRU configurations by varying the dimensionality of the GRU's hidden units. We observe that the relationship between PER and model size approximately follows a power-law-like trend similar to figure \ref{fig:perSize}. 

\subsection{\textsc{Data}$_{\textsc{ nato-words}}$}

In this dataset, {4} individuals silently articulated English sentences in a spelled-out format using NATO phonetic codewords. For example, the word \textsc{$<$rainbow$>$} was articulated as \textsc{$<$romeo-alfa-india-november-bravo-oscar-whiskey$>$}, with phonemic transcription \textsc{$<$r-ow-m-iy-ow {\scriptsize space} ae-l-f-ah {\scriptsize space} ih-n-d-iy-ah {\scriptsize space} n-ow-v-eh-m-b-er {\scriptsize space} b-r-aa-v-ow {\scriptsize space} ao-s-k-er {\scriptsize space} w-ih-s-k-iy$>$}. Subjects articulated the phonetically balanced \textsc{rainbow} and \textsc{grandfather} passages in this spelled-out format. In total, {1968} NATO codeword articulations were recorded across both passages, along with an additional {520} isolated codeword recordings used for training. EMG was recorded from {22} sites on the neck and cheek at a sampling rate of {5000}~Hz (electrodes were not placed on the right side of the neck; middle image in figure~\ref{fig:electrodePlacement}).

\begin{figure*}[htpb]
  \centering
  \begin{subfigure}[t]{0.48\textwidth}
    \includegraphics[width=0.8\linewidth]{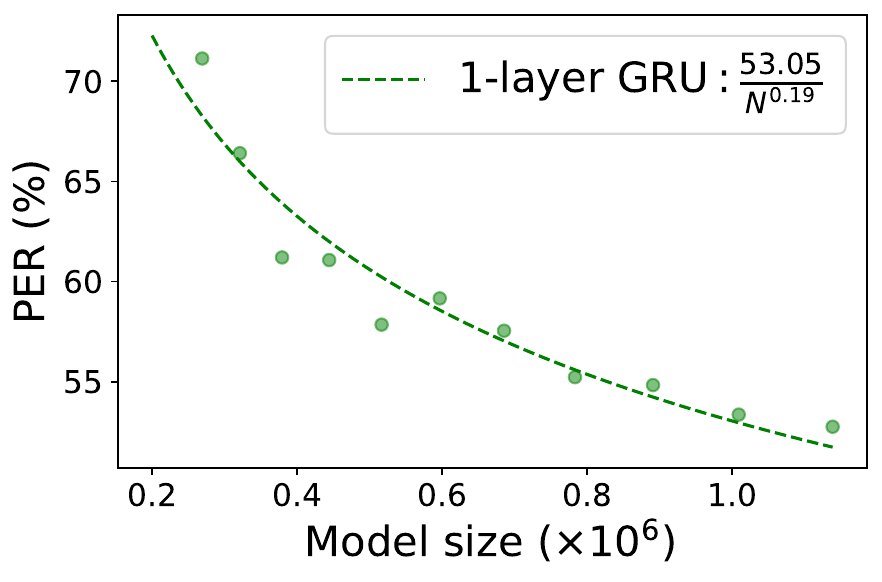}
    \caption{Subject 1}
  \end{subfigure}
  \hfill
  \begin{subfigure}[t]{0.48\textwidth}
    \includegraphics[width=0.8\linewidth]{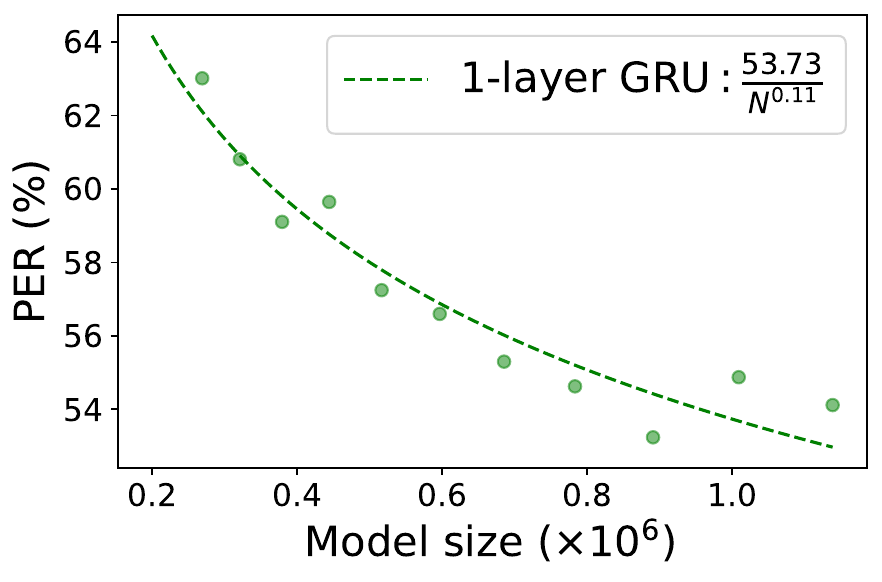}
    \caption{Subject 2}
  \end{subfigure}

  \vspace{1em}

  \begin{subfigure}[t]{0.48\textwidth}
    \includegraphics[width=0.8\linewidth]{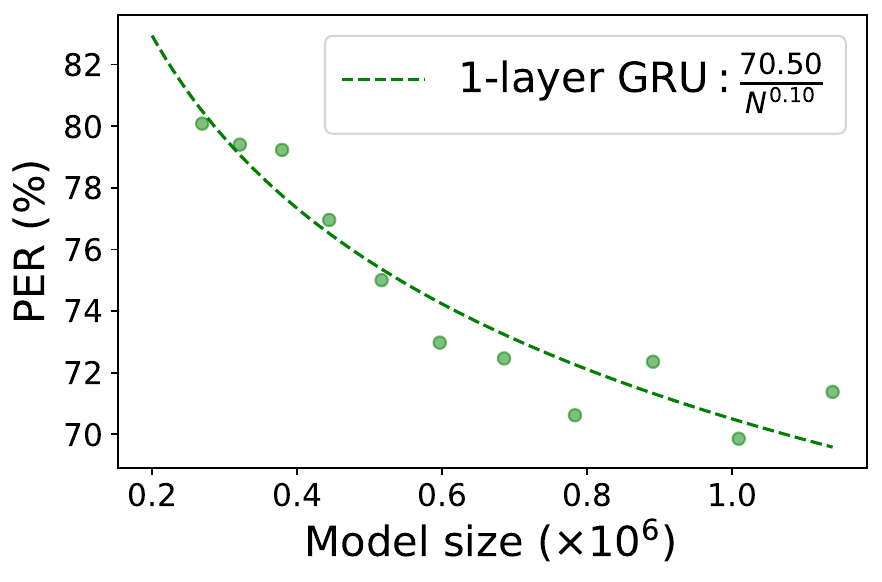}
    \caption{Subject 3}
  \end{subfigure}
  \hfill
  \begin{subfigure}[t]{0.48\textwidth}
    \includegraphics[width=0.8\linewidth]{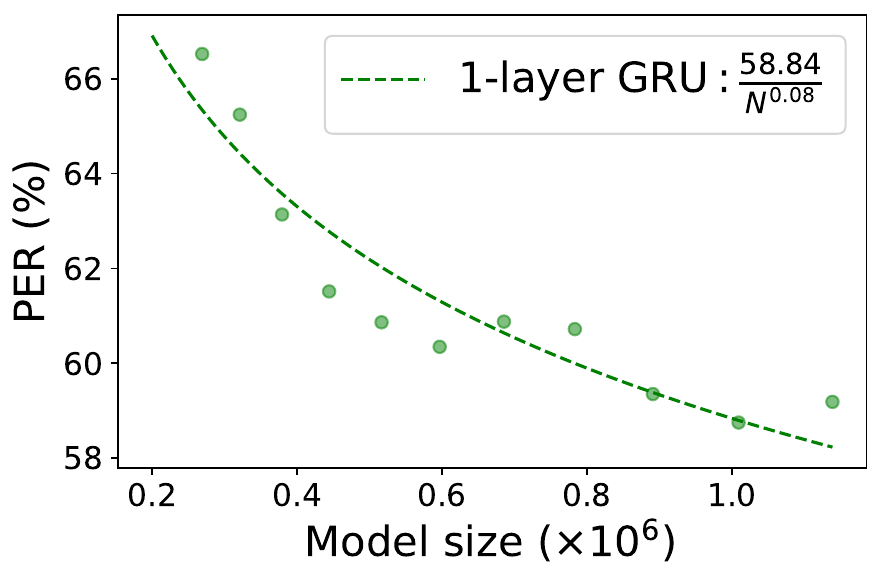}
    \caption{Subject 4}
  \end{subfigure}

  \caption{Model size versus PER for EMG-to-phoneme translation for \textsc{Data}$_\textsc{ nato-words}$.}
  \label{fig:natoWords}
\end{figure*}

\textbf{Results: } we use a timestep $\tau$ of 30 ms, implemented as a sliding window with 150 ms of overlapping context and a 30 ms step size, to compute $\mathcal{E}(\tau)$ and $\sigma(\tau)$, both of which are SPD matrices of size $22 \times 22$. The matrices $\sigma(\tau)$ are then input to a GRU for EMG-to-phoneme sequence translation. The dataset is split into training, validation, and test sets consisting of 416, 104, and 1968 articulations, respectively. The model depicted in figure \ref{fig:Illustration} is trained using a single GRU layer for 100 epochs, and the weights corresponding to the lowest validation loss are selected.

In table~\ref{tab:nato}, we report the character error rate (CER). For a given character articulation—for example, \textsc{$<$r$>$}, which corresponds to the spoken form \textsc{$<$romeo: r-ow-m-iy-ow$>$}—we consider the decoded character to be \textsc{$<$r$>$} if the predicted phoneme sequence most closely matches that of \textsc{$<$r$>$} among the 26 alphabet characters.
It is worth noting that the test set is nearly five times larger than the training set. This experimental paradigm is designed to evaluate whether a model can be trained effectively using very limited data—an important consideration for clinical applications, where collecting large amounts of data can be too strenuous for patients. In our case, the model is trained on just 10 minutes of data and evaluated on 50 minutes of data.

\begin{table}[h!]
\centering
\caption{Mean character error rate (CER) on \textsc{Data}$_{\textsc{ nato-words}}$. Lower CER indicates better performance. The chance-level CER is 96\%, and all subjects achieve substantially lower error rates.}
\label{tab:nato}
\begin{tabular}{p{1.5cm} p{2cm}}
\hline
\textsc{Subject} & CER (\footnotesize$\%\downarrow$) \\
\hline
1 & 55.7 \\
2 & 55.0 \\
3 & 70.4 \\
4 & 56.4 \\
\hline
\end{tabular}
\end{table}

In figure \ref{fig:natoWords}, we examine how model size across various GRU configurations affects the
PER. To do this, we vary the dimensionality of the GRU’s hidden units. We observe similar trends as noted in figure \ref{fig:perSize} across all subjects.

\subsection{Discussion}
These results indicate that accurate decoding is achievable in a small-vocabulary setting, suggesting that a minimum-viable neuromuscular speech prosthesis may be feasible even with limited training data.

\begin{table*}[b]
\centering

\caption{Examples of EMG-to-phoneme sequence translations. We do translations using EMG collected during {\em silent} articulations ($E_S$) with CTC loss without making use of corresponding time-aligned {\em audio} ($A$) and EMG collected during {\em audible} articulation ($E_A$). \textcolor{blue}{Ground truth sentences with corresponding timestamps.} \textcolor{green!30!black}{Ground truth phonemic transcriptions.} \textcolor{purple}{Decoded phonemic transcriptions.} \textcolor{orange}{Decoded sentences.}}
\begin{tabular}{l}
\hline
{Top-3 (best) transcribed sentences}\\
\hline
$_{\textsc{t-start}}<$\textcolor{blue}{\textsc{It Was Paid For}}$>_{\textsc{t-end}}$\\
\vspace{0.2cm}
\textcolor{green!30!black}{\textsc{ih-t $\textsc{\tiny space}$ w-aa-z $\textsc{\tiny space}$ p-ey-d $\textsc{\tiny space}$ f-ao-r}}\\
\textcolor{purple}{\textsc{ih-t $\textsc{\tiny space}$ w-aa-z $\textsc{\tiny space}$ p-ey-t $\textsc{\tiny space}$ f-ao-r}}\\
\textcolor{orange}{\textsc{It Was Pay For}}\\
\hline

$_{\textsc{t-start}}<$\textcolor{blue}{\textsc{It's A Community Center}}$>_{\textsc{t-end}}$\\
\vspace{0.2cm}
\textcolor{green!30!black}{\textsc{ih-t-s $\textsc{\tiny space}$ ah $\textsc{\tiny space}$ k-ah-m-y-uw-n-ah-t-iy $\textsc{\tiny space}$ s-eh-n-t-er}} \\
\textcolor{purple}{\textsc{ih-t-s $\textsc{\tiny space}$ ah $\textsc{\tiny space}$ k-ah-m-y-uw-n-ih-t-iy $\textsc{\tiny space}$ s-eh-n-t-er-n}}\\
\textcolor{orange}{\textsc{It's A Community Center}}\\
\hline

$_{\textsc{t-start}}<$\textcolor{blue}{\textsc{Just All Different Colors}}$>_{\textsc{t-end}}$\\
\vspace{0.2cm}
\textcolor{green!30!black}{\textsc{j-ah-s-t $\textsc{\tiny space}$ ao-l $\textsc{\tiny space}$ d-ih-f-er-ah-n-t $\textsc{\tiny space}$ k-ah-l-er-z}} \\
\textcolor{purple}{\textsc{j-ah-s-t $\textsc{\tiny space}$ ao-l $\textsc{\tiny space}$ d-ih-f-er-ah-n $\textsc{\tiny space}$ $\textsc{\tiny space}$ k-ih-l-er-z}}\\
\textcolor{orange}{\textsc{Just All Different Colors}}
\\
\hline
Bottom-3 (worst) transcribed sentences. \\
\hline

$_{\textsc{t-start}}<$\textcolor{blue}{\textsc{The Death Penalty}}$>_{\textsc{t-end}}$\\
\vspace{0.2cm}
\textcolor{green!30!black}{\textsc{dh-ah $\textsc{\tiny space}$ d-eh-th $\textsc{\tiny space}$ p-eh-n-ah-l-t-iy}} \\

\textcolor{purple}{\textsc{ih $\textsc{\tiny space}$ dh-ih-t $\textsc{\tiny space}$ ih-k $\textsc{\tiny space}$ p-ay $\textsc{\tiny space}$ ae-k}}\\
\textcolor{orange}{\textsc{That Thick My Back}}\\
\hline

$_{\textsc{t-start}}<$\textcolor{blue}{\textsc{He Does The Yard}}$>_{\textsc{t-end}}$\\
\vspace{0.2cm}
\textcolor{green!30!black}{\textsc{hh-iy $\textsc{\tiny space}$ d-ah-z $\textsc{\tiny space}$ dh-ah $\textsc{\tiny space}$ y-aa-r-d}}\\
\textcolor{purple}{\textsc{ih-ih-t $\textsc{\tiny space}$ ih-s $\textsc{\tiny space}$ n-ih-n-t $\textsc{\tiny space}$ ay-t}}\\
\textcolor{orange}{\textsc{It Its Knit Might}}\\
\hline

$_{\textsc{t-start}}<$\textcolor{blue}{\textsc{That's Awful}}$>_{\textsc{t-end}}$\\
\vspace{0.2cm}
\textcolor{green!30!black}{\textsc{th-ae-t-s $\textsc{\tiny space}$ aa-f-ah-l}}\\

\textcolor{gray}{\textsc{dh-eh-r $\textsc{\tiny space}$ ah $\textsc{\tiny space}$ t-oy-t}}\\
\textcolor{orange}{\textsc{There A Point}}\\
\hline
\end{tabular}
\label{tab:transcriptionEx}
\end{table*}
\end{document}